# Water at the molecular frontier

Mischa Bonn[1,*], Marie-Laure Bocquet[2], André Geim[3,4], Laura Fumagalli[3,4], Angelos Michaelides[5,*], Lydéric Bocquet[2,*],

[1] Department of Molecular Spectroscopy, Max Planck Institute for Polymer Research, Mainz 55128, Germany

[2] Laboratoire de Physique de l'École Normale Supérieure, Université PSL, Paris 75005, France

[3] Department of Physics and Astronomy, University of Manchester, Oxford Rd, M13 9PL, Manchester, UK

[4] National Graphene Institute, University of Manchester, Booth St E, Manchester M13 9PL, UK

[5] Yusuf Hamied Department of Chemistry, University of Cambridge, Cambridge CB2 1EW, United Kingdom; Lennard-Jones Centre, University of Cambridge, Cambridge CB2 1TN, United Kingdom

(*): bonn@mpip-mainz.mpg.de, am452@cam.ac.uk, lyderic.bocquet@ens.fr

**Abstract**

Much of the water that matters most — flowing through biological cells and membranes, percolating through soils and rocks, or processed by filtration and desalination technologies — is not bulk water. It is water in nano- and molecular-scale confinement or at interfaces. Nanoscale water is therefore central to many challenges of our time, from the global water crisis and the water–energy nexus to catalysis, electrochemistry, and health. Beyond its applications, nanoscale water has emerged as a rich scientific field in its own right: discoveries are accumulating rapidly, revealing that confinement and interfacial forces produce a range of behaviours absent in the bulk. Sometimes this amounts to a renormalization of bulk properties; in other cases, new and non-intuitive phenomena emerge — exotic phases, anomalous transport, and dielectric responses that defy continuum expectations. A convergence of atomically precise nanofluidic devices, novel spectroscopies and microscopy approaches, and machine-learning-accelerated simulations is now exposing the molecular mechanisms behind these behaviours. This Perspective argues that the recent findings should not be viewed as isolated nanoscale anomalies, but as evidence for a molecular regime of water in which continuum descriptions must be rethought and molecular-level understanding becomes essential. Recognising and exploring this regime should have consequences ranging from fundamental physics to the rational design of future technologies that rely on water and dissolved ions.

**Water's molecular moment**

Water underpins life, civilization, and the global economy, yet it is increasingly scarce, stressed, and contested. A 2026 United Nations report declared an era of "global water bankruptcy"[1]: chronic over-abstraction, pollution, and climate-driven shifts in the hydrological cycle are depleting groundwater reserves that took millennia to accumulate, and roughly half the global population already experiences water scarcity. Addressing this crisis demands not only better governance and infrastructure but, fundamentally, better technology — and that in turn requires a deeper molecular understanding of water itself.

A large fraction of technologically and biologically relevant water is either interfacial or confined: in desalination and fuel-cell membranes, in geological nanopores, in battery electrolytes, and in the hydration shells of proteins and membranes. In all these settings, surface-to-volume ratios are high and interfacial forces compete with, or overwhelm, bulk-like hydrogen bonding, so that structure, dynamics, and chemistry become tightly coupled. Nanoscale water, therefore, sits at the heart of the water-energy nexus — desalination, blue-energy harvesting from salinity gradients, and electrochemical technologies for hydrogen production and storage — as well as in catalysis, health, and separations more broadly. In many of these areas, the limiting factor is our ability to predict and control water at the molecular level.

Encouragingly, nanoscale water is a rapidly maturing scientific field. Over roughly the last decade, three developments have transformed the field from a collection of intriguing anomalies into a landscape where predictive understanding is within reach. First, low-dimensional materials and advanced nanofabrication now enable nanofluidic devices with atomically smooth channels and well-defined separations, ranging from single- or multi-walled carbon or boron nitride (BN) nanotubes[2–5] to nanoslits assembled from 2D materials[6,7]. Second, a new generation of spectroscopic and microscopic tools — including single photon emitters[8,9], sum-frequency generation (SFG)[10–12], ultrafast vibrational and dielectric spectroscopies[13,14], quantum sensing with nitrogen-vacancy centers[15,12,16], and related approaches[17] — can probe interfacial and confined water with molecular specificity, from hydrogen-bond networks to local dielectric response. Third, machine-learning (ML) interatomic potentials[18,19] provide quantum-accurate simulations of bulk, interfacial, and confined water over longer length and time scales, turning earlier qualitative atomistic models and their observables into quantitative predictions.

The power of this toolbox lies less in any single technique than in their convergence. Nanofluidic transport measurements reveal flow rates, slip lengths, and ionic conductivities; spectroscopies report on orientational structure, hydrogen-bond dynamics, diffusivity, and dielectric response; and ML-accelerated molecular dynamics links these observables to microscopic configurations and rare events. Together, they are beginning to clarify what nanoscale confinement does to water, and why it matters for membranes, electrochemical devices, and separation technologies.

This is also what makes nanoscale water such a compelling and fertile scientific frontier. The central question is no longer simply whether confinement changes water, but how those changes should be described. In this context, continuum models treat water as a smooth medium described by a few macroscopic fields and parameters — density, viscosity, a scalar dielectric constant, equations such as Navier–Stokes, Kelvin's, Poisson-Boltzmann, or Poisson–Nernst–Planck — without considering individual molecules or hydrogen bonds. In some cases, confinement and interfaces merely renormalise familiar bulk concepts. Renormalized continuum models retain this framework by allowing those effective parameters and boundary conditions to depend on confinement and interfacial chemistry: slip lengths become large, dielectric constants become anisotropic or geometry-dependent, and interfacial layers are represented by modified "effective" properties. In contrast, true molecular-level anomalies are situations where no plausible renormalization of bulk parameters within a homogeneous continuum can capture the observations, because the physics hinges on discrete layering, specific hydrogen-bond arrangements, strong correlations, or quantum effects — for example, the emergence of low-dimensional ice phases, giant dielectric anisotropy, or correlated ionic transport exhibiting Coulomb blockade or memristive (history dependent) behaviour. For these cases, the very language of homogeneous continuum water fails. Molecular-scale water is therefore a natural arena in which to ask where continuum descriptions survive, where they fail, where they bend, and then how molecular-level insight can be translated into predictive laws for technology.

In this Perspective, we provide a brief - and rather personal - take on the rapidly evolving frontier of molecular-scale water, exploring it across three overlapping domains: structural and dielectric properties, transport and friction, and technological implications. For complementary views and reviews of the topic, see refs.[20–22,5,23–25].

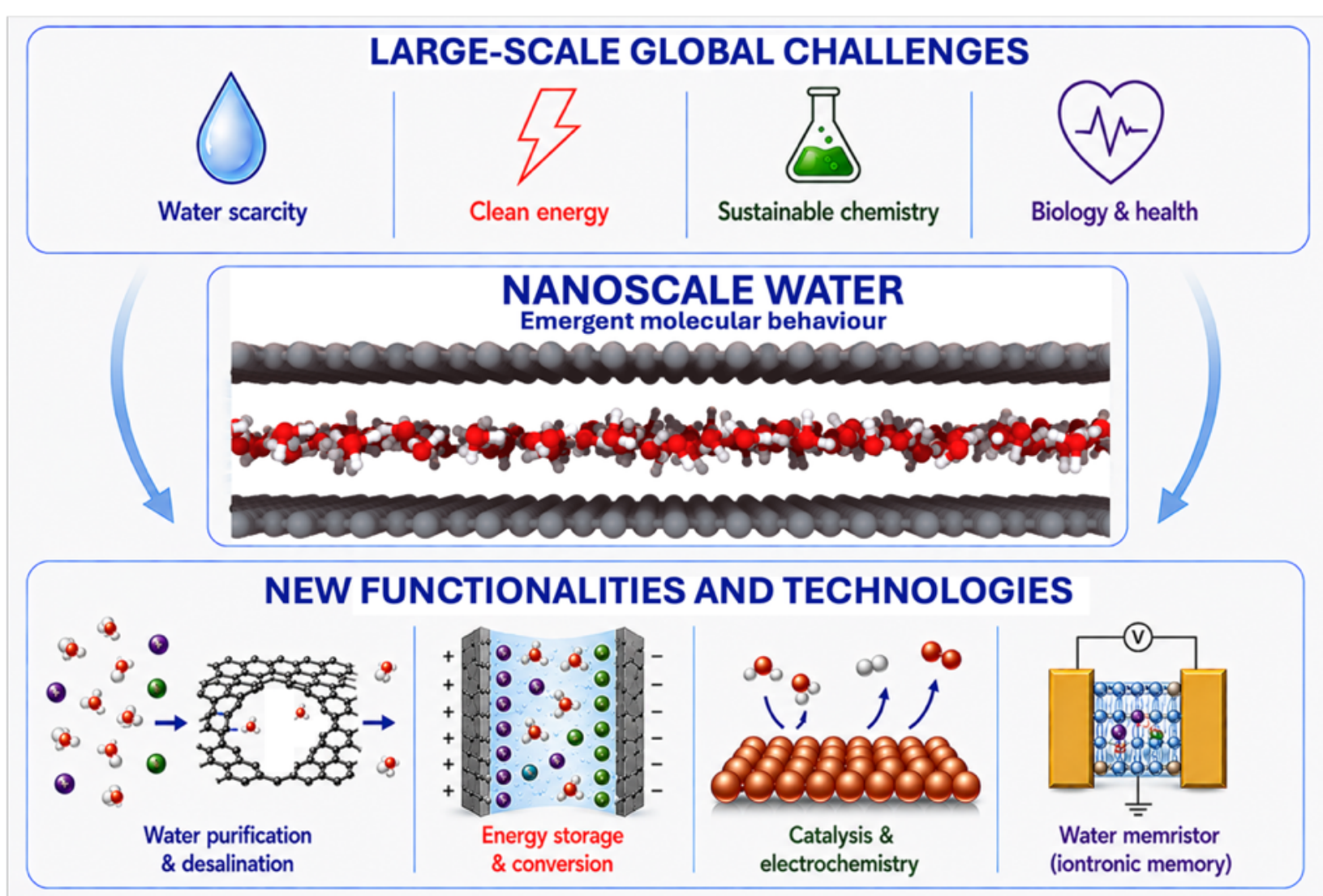

*Figure 1. Several important water-related technologies rely on the molecular properties of nanoscale water, including water purification, desalination, selective ion transport, and hydrogen generation from water. Despite the apparent technological impact of understanding nanoscale water, many fundamental insights are lacking. A fundamental, molecular-level understanding of nanoscale water is expected to enable novel, clever technologies beyond the current capabilities. We argue that the bottleneck for new technologies is our limited understanding of nanoscale water.*

## Structure, response, reactivity

Confinement reorganises water at the molecular scale, with consequences for dielectric response, electrostatic screening, chemical reactivity, etc. This raises the question of when bulk concepts such as phase diagrams, Kelvin's equation, and a scalar dielectric constant of $\varepsilon \approx 80$ can be rescued by renormalised parameters, and when the very notion of a homogeneous continuum becomes untenable.

**The continuum is dead, long live the continuum**

Bulk water has a peculiar and well characterised phase diagram[26]. In nanoconfinement, direct structural characterisation is more challenging because the relevant interfaces are buried. Recently, atomically defined carbon-based systems have provided important model platforms. High-resolution electron microscopy of few-layer water confined between graphene sheets suggested square-like ice at room temperature[27], highlighting the role of lateral confinement pressures in stabilising dense monolayer phases, even if the structural assignment remains debated[28]. ML-accelerated simulations of water in few-molecule-wide graphene-like slits predict temperature–pressure phase diagrams with an impressive diversity of phases, including hexagonal, square, rhombic, pentagonal, hexatic, and superionic states[29,30] (Fig. 2A) These studies emphasize that even "simple" slit confinement fundamentally reshapes the structural landscape of water by introducing high pressures[27] and altering orientational degrees of freedom.

Spectroscopic probes complement these structural images by revealing how confinement modifies hydrogen bonding at room temperature. Heterodyne-detected SFG (HD-SFG) of water confined between graphene and $CaF_2$ reveals that water behaves essentially as two independent interfacial layers plus a bulk-like interior until the separation decreases to about three molecular layers[10,31] (Fig. 2B). Only when the confinement becomes truly angstrom-scale, do signatures of weakened hydrogen bonding and orientational disorder appear, which cannot be described as a simple superposition of independent interfaces. This crossover from interface-dominated to confinement-dominated behaviour at approximately three layers, also seen in simulations[32], provides a useful rule-of-thumb: "nanoconfined" water in many real pores is structurally more akin to bulk water with modified boundary layers, rather than fully reconstructed multilayers. Importantly, this crossover at the nanoscale also matches the limit of validity of the Navier-Stokes equation for water[33], and coincides with pronounced changes in dielectric response[13,14].

These observations provide a nuanced view of continuum breakdown at the molecular scale. On the one hand, nanofluidic flow in many systems can be rationalised by continuum hydrodynamics with renormalised boundary conditions down to a few-nanometre scales, as long as one accounts for slip, modified viscosity, and coupling to ion transport. On the other hand, the same systems may host structurally exotic phases at higher lateral pressures or in narrower channels, and these phases feedback on transport by altering friction, dielectric response, and reactivity. Therefore, there is no single "nanoscale and molecular-scale water" regime.

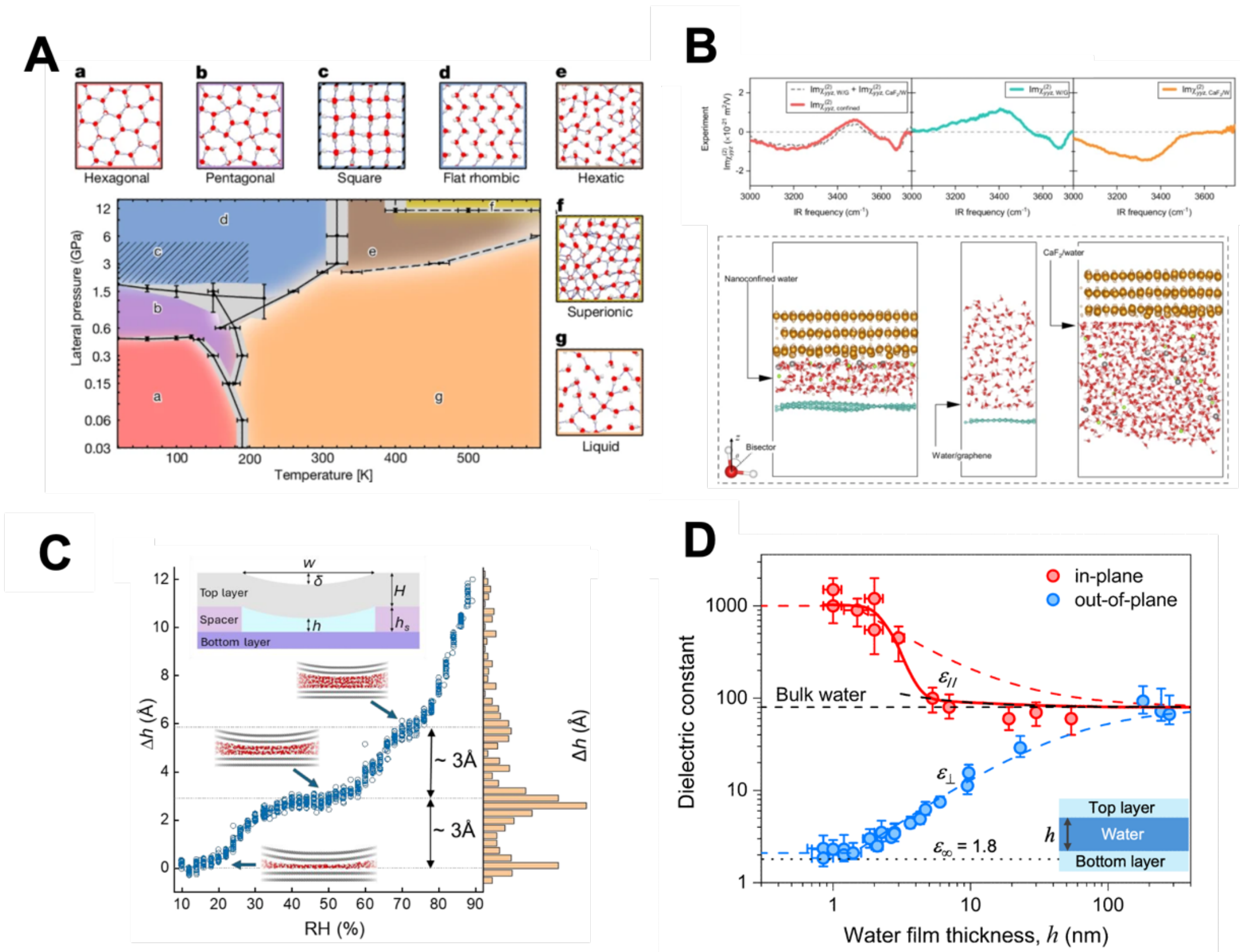

***Figure 2. A cabinet of curiosities for the structure and dielectric response of nanoconfined water.*** *A. Simulations predict rich phase behaviour for monolayer water*[29]*. B. SFG of nanoconfined water in a graphene/*$CaF_2$ *slit reveals that only below about 1 nm are genuine nanoconfinement effects observed*[10]*. C. Capillary condensation can happen in a step-wise (layer-by-layer) manner*[34]*. D. Both the in-plane and out-of-plane dielectric constants of nanoconfined water can deviate significantly from the bulk value*[13,14].

**Kelvin in disguise**

Capillary condensation is usually regarded as a well-understood macroscopic phenomenon, one that governs how water fills the smallest gaps in solids, rocks, powders, and microelectronic devices as well as those between contacting surfaces[35,36]. Yet its underlying physics often lies at the molecular scale. Under ambient humidity, the relevant gaps are only about one nanometre - a few water molecules across. The resulting nanoscale water bridges that form between nearby surfaces control adhesion, stiction, friction, lubrication, corrosion, and cohesion, see e.g.[37]. Condensation is therefore one of the most ubiquitous situations in which the molecular discreteness of water might be expected to challenge continuum thermodynamics. Nonetheless, the century-old Kelvin equation remains the workhorse of capillary condensation, even though it treats water as a continuous liquid.

Why does the continuum picture often survive, and where does it fail? The Kelvin equation relates the condensation threshold to the relative humidity and the properties of the confining material.

Part of the Kelvin equation's apparent success stems from the use of effective contact angles, wetting-film thicknesses or pore sizes as adjustable parameters[38]. Recent experiments using capillaries assembled with angstrom-scale precision have provided a clearer, although still emerging, perspective[34,38]. Under molecular-scale confinement, water forms discrete layers, so that its free energy oscillates with layer number. Rigid pores cannot adjust appreciably towards favourable integer-layer configurations. Commensurability – the "fitting" of discrete water layers within the pores – can then markedly shift the condensation transition relative to the Kelvin prediction, favouring commensurate pores over incommensurate ones. Even so, the transition manifests itself experimentally as an abrupt change, resembling the Kelvin continuum picture. The molecular discreteness thus affects water condensation by shifting its threshold, yet is hidden from direct view.

Wall compliance changes this picture further. Capillary pressures in nanoscale pores can reach kilobar values, large enough to deform sufficiently compliant walls and allow pores to relax towards commensurate configurations accommodating an integer number of molecular layers. Such elastic adjustment can partly smooth the oscillatory contribution to the free energy, helping Kelvin-like behaviour persist down to the few-molecular-layer limit[38]. If the confining walls are compliant enough to adjust on the angstrom scale, the discreteness becomes directly apparent (Fig. 2C): water filling can proceed layer by layer, in steps of approximately one molecular diameter[34]. Wall compliance is therefore not a minor correction but a control parameter.

Beyond structure, this highlights a broader principle for water at the nanoscale and in membranes: the compliance of confining walls and their intrinsic thermal fluctuations can reshape the structure and dynamics of confined water, thereby fundamentally modifying the fluid response[39–42].

**Water forgets its 80**

Structure alone does not capture the peculiarities of molecular-scale water. Equally important is how water responds to electric fields, because this governs electrostatic screening, solvation, ion transport, and chemical reactivity. Bulk liquid water is already exceptional in this respect: its large dielectric constant, $\varepsilon \approx 80$, reflects collective polarization of the hydrogen-bond network, not just the response of individual molecular dipoles. At interfaces and under strong confinement, however, this familiar number becomes an inadequate descriptor. Water does not merely acquire a different value of $\varepsilon$: interfaces disrupt and reorganise the hydrogen-bond network, making the electrical response highly anisotropic and spatially nonuniform.

The expectation that interfacial water should lose its bulk-like $\varepsilon$ is nearly a century old[43–45]. However, demonstrating this experimentally was challenging and debated[46–49]. Addressing this challenge became possible only recently through the application of nanoscale dielectric spectroscopy to slit-like nanochannels assembled from atomically flat crystals[13]. Measurements of individual water-filled channels showed that water films thinner than ~2 nm behaved almost like electrically dead layers. Their out-of-plane dielectric response was nearly completely suppressed, with $\varepsilon_\perp$ falling to ~2 (Fig. 2D). The suppression remained substantial even in confined water thicker than 10 nm because the effective perpendicular response was dominated by ultra-low-$\varepsilon$ interfacial layers. Several optical, spectroscopic, and transport studies are broadly consistent with this conclusion, although they probed different frequencies, electric-field directions, and interfacial environments[50–54].

The in-plane response of nanoconfined water is also anomalous, but in a different way. It remains bulk-like up to gigahertz frequencies in water layers thicker than ~3 nm (Fig. 2D), but in thinner layers, it increases to very large, ferroelectric-like values, with $\varepsilon_\parallel$ reaching ~1,000[14]. Confinement, therefore, turns molecular-scale water into a strongly anisotropic medium where ε can no longer be treated as a scalar: polarization perpendicular to surfaces is quenched, while in-plane polarization remains large and may even be enhanced.

The same measurement platform also showed that water near the confining walls exhibited high in-plane conductivity, ~0.1 S $m^{-1}$ (Fig. 3D, red symbols), consistent with earlier studies of interfacial water[55–57]. Under stronger confinement, the conductivity rose sharply, reaching several S $m^{-1}$ at thicknesses of ~1.5 nm[14]. This superionic-like behaviour coincided with the anomalous dielectric response at similar thicknesses, suggesting that polarization and conduction are two faces of the same correlated dynamics. For still thinner water layers, conductivity decreased again, consistent with transport experiments on bilayer and monolayer water (Fig. 3D, grey symbols), where extreme confinement may limit fast Grotthuss-type proton transport[58,59].

Simulations reproduce, at least qualitatively, the strong dielectric anisotropy of interfacial and confined water[60–65]. Analytical approaches have made further progress by linking the suppressed out-of-plane response to molecular correlations and structural ordering at interfaces[66,67], and reproducing experimental trends quantitatively[67]. Yet a unified framework is still missing that connects $\varepsilon_\perp$, $\varepsilon_\parallel$, and the enhanced in-plane conductivity on the same molecular footing.

Overall, these developments change how molecular-scale water should be described. Continuum electrostatic and electrokinetic frameworks may remain useful, with quantities such as Debye and Bjerrum lengths still serving as effective concepts, but their usual premise - that water can be represented by the simple number $\varepsilon \approx 80$ – clearly breaks down near surfaces. Instead, ε becomes a tensor varying over angstrom distances, and its perpendicular and parallel components can differ by orders of magnitude. Most direct measurements have focused on atomically flat, chemically simple interfaces. Realistic systems introduce surface charge, roughness, curvature, compliance and dissolved ions, all of which can reshape the dielectric response. Extending this emerging picture to such environments, and to quasi-one-dimensional water, is now the challenge for experiment and theory.

These dielectric anomalies and interfacial reorganisations are closely linked to chemistry, from autoprotolysis in confined water to spontaneous surface charging in 2D-material interfaces, as discussed next.

**Water at the reactive edge**

Self-dissociation of water into hydronium and hydroxide is fundamental to aqueous acid–base chemistry and much of electrochemistry. Even in bulk, it is a delicate balance controlled by covalent bond breaking, charge transfer, hydrogen-bond fluctuations, and nuclear quantum effects. At reactive interfaces and under confinement, this balance can shift markedly, altering both water's autoprotolysis and its chemical reactivity. On metals such as Ru, strong bonding of hydroxyls and hydrogen to the surface stabilises dissociated configurations so that a substantial fraction of interfacial water molecules are effectively split, whereas on Pt or Pd interfacial water largely remains molecular, despite adopting a flat adsorption geometry[68,69]. 2D materials like pristine graphene and hBN show no evidence of dissociation of the complete first adsorbed water layer. However, surface defects[8,70] can induce water dissociation, as revealed by *ab initio* molecular dynamics (AIMD) simulations. Even at nominally defect-free hBN–water interfaces, experiments revealed that contact with water induced a substantial negative surface charge at near-neutral pH, in contrast to graphene. pH-dependent measurements and simulations identify hydroxide ions as the main cause of charging[10,71].

In nanoconfined geometries between 2D materials, the picture is more nuanced. Confinement can generate high effective pressures and modify coordination environments, which in turn alter hydrogen-bond networks and electrostatic screening, both of which are key determinants of dissociation thermodynamics. Yet direct experimental access to self-ionization in confined water remains challenging, so that much of our insight comes from simulations that reach conflicting conclusions: some report enhanced dissociation, others suppressed or nearly unchanged reactivity relative to bulk[72–74]. Recent studies using ML potentials trained on density-functional theory data offer a unifying perspective[75], showing that most of the apparently exotic reactivity can be rationalised as the ordinary response of water to unusual thermodynamic conditions. The high effective pressure reached in nanoconfinement provides a powerful way of accessing otherwise inaccessible thermodynamic states, rather than a source of fundamentally new chemistry. In the ultimate confinement of monolayer water, AIMD simulations complemented by ML molecular dynamics do hint at modifications in surface reactivity, including the demonstration that $OH^-$ adsorption depends on stacking between hBN walls[76] [Sanguinetti2026]. Such effects vanish in bilayer water that largely behaves as two decoupled systems consistent with SFG measurements for graphene and $CaF_2$ walls discussed earlier[10].

While confinement can often be viewed as pushing water into unusual thermodynamic states, chemically heterogeneous walls, defects that localise charge, extreme curvature and coupled electronic degrees of freedom may create local environments in which reaction pathways and barriers differ qualitatively from those in bulk water. The challenge is to test this directly by combining experiments and simulations.

## Transport and friction

Once water's structure and dielectric response are altered, its flow, friction and charge transport inevitably change too. This section asks how far classical hydrodynamics and electrokinetics can be pushed to describe nanoscale water using modified boundary conditions, and when molecular correlations, ionic discreteness and couplings to the confining solid force a description beyond continuum theories.

**Big changes in tight spaces**

Molecular-scale confinement creates tight feedback between structure and transport. Small changes in local hydrogen-bond topology and interfacial ordering can produce orders-of-magnitude variations in water and ion mobility, friction, and slip length. This parallels established findings in biological channels, where transport efficiency is heavily dependent on exact structural configurations[77,78].

The interplay between structure and transport is perhaps most vividly illustrated by one-dimensional proton and ion conduction in subnanometre channels. In single-file water wires, as found in biological aquaporins or model nanotubes, proton transport proceeds by concerted hydrogen-bond rearrangements rather than independent hopping[79]. In single-file CNT channels, proton transport rates exceeded those in bulk water by an order of magnitude[80] (Fig. 3A). In contrast, in monolayer 2D water at ambient pressures, proton transport was anomalously slow even relative to the 3D case[59] (Fig. 3C), but increased by orders of magnitude in slightly thicker water layers, approaching superionic conductivity[14] (Fig. 3D) Simulations also show that at high pressures monolayer water is superionic[29,81] (Fig. 3B). This illustrates that proton conductivity is dramatically affected by subtle changes in confinement conditions.

Another classic example is water flow in carbon nanotubes, the "drosophila of nanofluidics". Continuum Navier-Stokes describes flow down to the nanometre scale, but experimental surface slip lengths increase dramatically as tube radius decreases below tens of nanometres[4,5,82–84]. Nanoscale slits with atomically smooth graphitic walls show a related effect: for channel heights below ~2 nm, water flow is strongly enhanced by the combination of low water–graphene friction and high capillary/disjoining pressures[85,6]. Changing the wall material adds another layer of sensitivity: slippage in BN nanotubes is considerably reduced compared to carbon nanotubes, even though the two exhibit nearly isostructural surfaces, a difference believed to arise mainly from spontaneous surface charging and hydroxide adsorption on hBN[3,70,86] (Fig. 3E).

These observations have driven a conceptual shift away from purely steric or real-space pictures of friction toward a view based on many-body couplings between collective modes of water and those of the confining solid. Structure factors in Fourier space then become an appropriate descriptor for transport[87]. In this framework, friction reflects the overlap of the liquid's low-frequency dielectric and vibrational spectrum—dominated by Debye relaxation and hydrogen-bond librations—with the phonon and electronic excitations of the wall. When these spectral densities resonate in both frequency and wavevector space, fluctuation-induced forces efficiently transfer momentum between water and the solid, increasing friction; if off-resonant, slippage is enhanced[88,89]. This mechanism provides a rationalisation for the radius-dependent slippage in multiwall carbon nanotubes (Fig. 3E).

ML-accelerated simulations, combined with spectroscopy and transport measurements, now link descriptors such as hydrogen-bond network connectivity, orientation distributions, and local dielectric response to macroscopic observables such as friction coefficients and ionic mobilities. Developing compact, transferable descriptors that bridge this structural–transport hierarchy is a challenge, but one that is central to both predictive modelling and materials design.

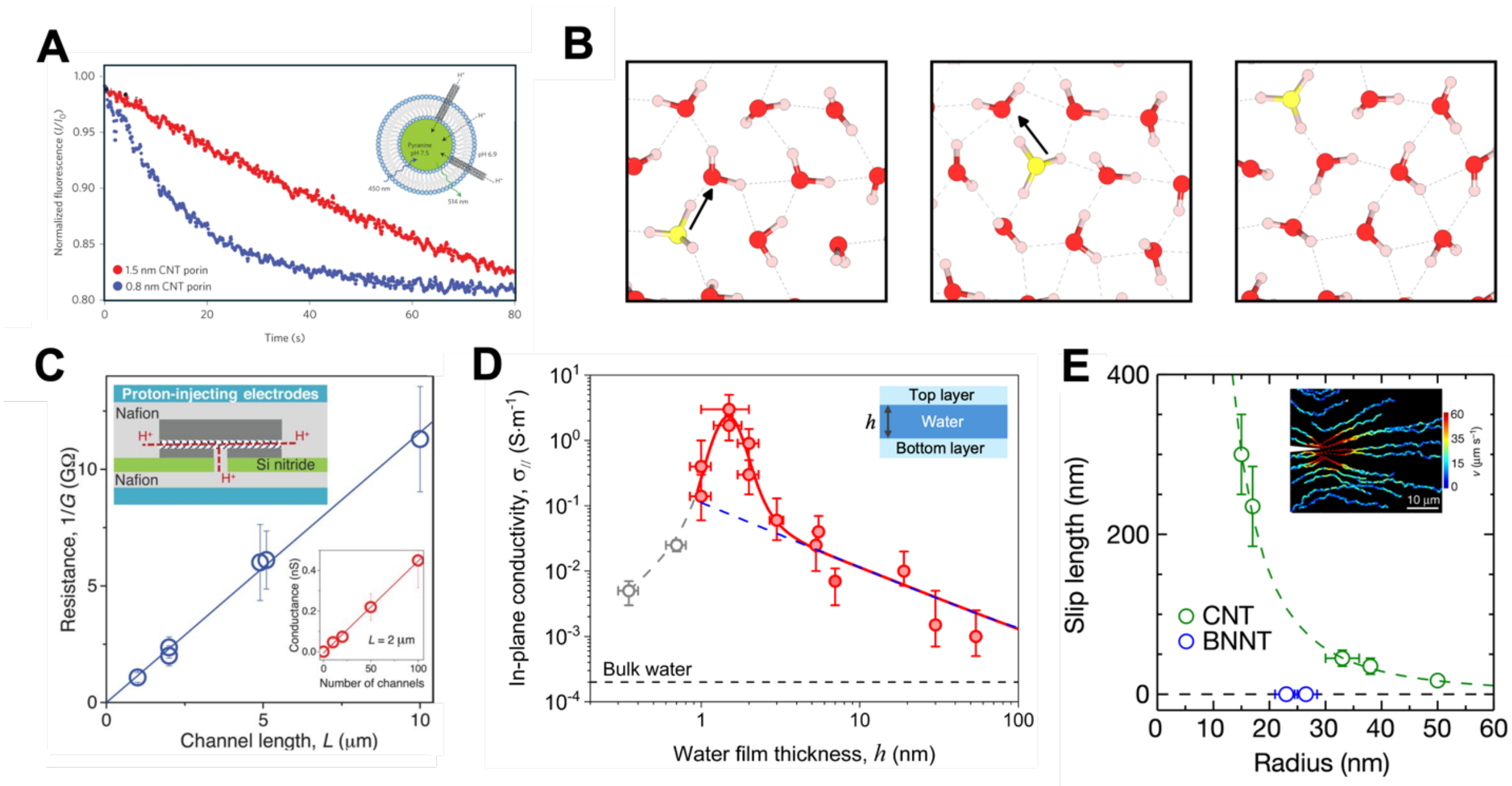

***Figure 3. Big transport changes in small spaces.*** *A. Evidence for rapid proton conduction in carbon nanotube channels*[80]*. B. Illustration of the Grotthuss mechanism of proton transport driving proton transport in (superionic) water*[81]*. C. Exquisitely precise nanochannels can probe proton transfer rates in a single layer of nanoconfined water*[59]*. D. High in-plane ionic conductivity of nanoconfined water*[14]*. E. Water flows rapidly through carbon nanotubes but not through hBN nanotubes*[3]*.*

## Beyond the average electrolyte

Electrolytes in confinement are both complex and functionally rich because the characteristic length scales of electrolyte theories often match the pore size. As a result, screening, ion correlations, and electrokinetic transport become intimately coupled to the dielectric and electronic properties of the confining material. Consequently, even a simple question such as how much salt enters a nanopore depends on the interplay of steric exclusion, ion-specific solvation, dielectric image charges, surface-charge-induced Donnan potentials, and ion–ion correlations[90,91].

Small changes in salt identity, surface chemistry, pore geometry, or thermodynamic conditions can produce qualitatively different outcomes: from strong exclusion and nearly deionised pores, to charge inversion and overscreened double layers, to regimes where individual ions enter or leave in discrete steps reminiscent of electronic Coulomb blockade[20,92–94] . This sensitivity is not a nuisance but an opportunity: it can be explored for highly selective separation, ionic gating, and nanofluidic devices. Nanoscale water forces us to abandon the comfort of universality and embrace singularity and diversity.

From a transport perspective, confined electrolytes expose both the strengths and limitations of classical electrokinetic frameworks. On the one hand, Poisson-Nernst-Planck & Stokes formalisms augmented with realistic surface charge distributions can successfully describe complex transport behaviour, such as rectification in asymmetric nanopores, mechano-sensitive conductance, and nonlinear electroosmotic flows[95]. On the other hand, some phenomena observed in strongly confined water cannot be accounted for by mean-field frameworks, and a proper rationalisation requires the inclusion of molecular or even electronic degrees of freedom[20,87]. For example, ionic correlations and pairing dynamics are a prerequisite to explain voltage-gating and ionic Coulomb blockade[96,97], as well as history-dependent conductance in memristive 2D slits[97,98] and the apparent breakdown of the Nernst–Einstein relation in carbon-based nanochannels is rationalised by water fragmentation under an electric field[99].

## The quantum life of water

Quantum mechanics underpins all aspects of water, but in nanoscale systems the question is which quantum effects go beyond the "usual" role in bonding and spectroscopy and become essential for understanding structure and transport. Nuclear quantum effects—zero-point motion and tunnelling— are at the core of isotopic effects[100,101]. Such effects can now be included with path-integral molecular dynamics, and these simulations show that quantum fluctuations enhance hydrogen-bond rearrangements and transient autoprotolysis events without dramatically altering bulk densities because of competing, near-cancelling quantum contributions[101]. At interfaces and in strongly confined phases, especially where water molecules are partially dissociated or constrained, such nuclear quantum contributions can be more pronounced, affecting both structure and proton transport[30,102,103].

More unconventional are phenomena in which the motion of water and/or ions couples to the electronic degrees of freedom of the confining solid through long-range Coulomb interactions. This effect, known as "quantum friction", has been suggested to underlie anomalous slippage of water inside multiwall carbon nanotubes discussed above. Further signatures of such electronic couplings were reported in the sensitivity of slippage to the metallicity of the confining tube[104], as well as light-induced effects on friction[105] and electron cooling via Terahertz spectroscopy at graphene–water interfaces[106]. Another hint comes from measurements of electronic currents generated by water flowing or by droplets moving along graphitic surfaces[107–110]. While ionic Coulomb drag is often invoked to explain these phenomena, recent theories predict that charge fluctuations in flowing water can directly exchange momentum with electrons in nearby graphene or other conductors, generating electronic currents even in the absence of net ionic flow[87].

These effects can all be rationalised within theoretical frameworks that treat friction as arising from overlap of liquid and solid fluctuation spectra[87]. Then, the key design parameter for "quantum-engineered" hydrodynamics is the spectral density of substrate modes, which can be tuned via material choice, doping, and nanostructuring. Understanding how these quantum-origin couplings interplay with classical hydrodynamics and with confinement-induced changes in Debye relaxation and hydrogen-bond dynamics is a promising direction for future work, with predictions ranging from flow tunnelling[111] and hydroelectronic energy conversion[112] to species separation based on their molecular signatures.

**The substrate strikes back**

As discussed above, the solid walls in nanoscale-water systems are not merely inert boundaries but can become active participants. The influence also runs in reverse. Increasing evidence shows that wetting can substantially modify the structure, chemistry, and excitations of the contacting surfaces. For example, many oxide and nitride surfaces reconstruct or hydroxylate when wet, altering local chemistry, phonon spectra, and mechanical properties[17,113].

These water-induced reconstructions modify how energy flows across interfaces. Changes in surface phonon modes and their lifetimes alter the density of states available for coupling to liquid vibrational modes[114,115], thereby tuning interfacial thermal conductance and vibrational energy relaxation pathways. Similarly, adsorbed water and ions change carrier densities in 2D materials such as graphene[116–118]. In systems where hydrodynamic Coulomb drag is significant, the resulting quantum feedback can reduce or enhance friction depending on how electronic and vibrational spectra overlap[87].

Recognising confined water as an active agent that co-evolves with its substrate has practical implications. For fundamental studies, it means that interpreting spectroscopic and transport measurements requires considering the coupled water–solid system rather than assuming unmodified, vacuum-characterised surfaces. For applications, it suggests that preconditioning with water or deliberate water-induced reconstruction could serve as a design tool—for example, to create catalytically active hydroxylated surfaces or to tune friction and thermal transport in membranes and heat exchangers. Challenges here are to develop models that capture these slow, often pH-dependent co-evolution processes alongside fast molecular dynamics and complex non-adiabatic couplings of electronic and nuclear degrees of freedom.

## From pores to technologies

Ultimately, any continuum–molecular reconciliation should prove its worth in practice. Biology already shows how molecular-scale confinement of water and ions can be harnessed for selectivity, gating and energy conversion, and synthetic nanofluidic devices are beginning to explore the same principles. The challenge is to turn molecular insight from model pores and individual nanodevices into functional technologies and, eventually, water infrastructure.

**Pores before engineers**

Biological systems demonstrate that water in angstrom- to nanometre-scale confinement can be exploited with exquisite precision. Aquaporins, with hourglass-shaped channels about 3 Å wide at their narrowest point, achieve almost perfect selectivity for water over ions while maintaining high permeabilities. They do this by choreographing single-file water wires and orientational frustration that blocks proton leakage[77,119]. Mechanosensitive channels such as Piezo or TRAAK transduce deformations into changes in ion and water flow, underpinning sensations like touch and hearing[120]. Proton pumps convert electrochemical gradients into rotary motion in bacterial flagellar motors, illustrating that water-mediated ion transport can be harnessed to produce mechanical work[121].

Synthetic nanofluidic devices need not replicate biological channels atom by atom. Instead, they can adopt the same operating principles to control water and ion transport. For example, whereas pore deformation underlies mechanosensitive conductance in biological channels[122], experiments on carbon nanotubes and 2D channels[123,124] have already demonstrated pressure-sensitive ionic transport without mechanical deformation. Instead, the response arises from nonlinear ion transport that is highly sensitive to surface charge patterns and interfacial slip. Memristive behaviour—history-dependent conductance—has been observed in many other nanoscale systems where ion dynamics produce long-lived states reminiscent of synaptic plasticity[25,98]. These examples show that nanofluidics can already emulate some basic functions of biological channels — gating, pumping and memory — using geometry, surface chemistry and the nonlinear response of confined electrolytes.

**Dreaming with ions**

Our brain is a massive ionic computer. Neural activity relies on subtle ionic balances and transport in nanopores, using ions as carriers for computation or memory[126]. An appealing long-term goal of nanofluidics is to build synthetic "ionic brains" in which information is stored and processed by ions in water rather than electrons in solids and to do computation with such devices. Ions are 'multiflavour' species which offer multiple internal degrees of freedom—valence, size, hydration structure—that could be exploited for richer logic than binary bit states, at the cost of slower speeds. Recent breakthroughs in neuromorphic nanofluidics indicate a possible route to such ionic computation[25,98,127]. Studies have demonstrated manifold memristive behaviours in confined electrolytes[25], harnessing various nanoscale assets, from ionic correlations to dynamical surface adsorption. This rapid progress has already yielded elementary processing capabilities, including Hebbian learning and reservoir computing. However, realising neuromorphic computation requires the ability to construct large-scale architectures composed of numerous interacting nanofluidic elements. Building such architectures remains a huge, largely unexplored challenge. A more immediate goal is to identify where ion-tronic processing can offer genuine advantages over electronics, particularly in aqueous, soft or bio-inspired environments that naturally couple memory, sensing and transport.

**Nanopores by the acre**

Much of the excitement around nanoscale water arises from spectacular behaviours in single nanochannels. The question for the water–energy nexus is whether such phenomena can make a difference at the scale of rivers, desalination plants, and power grids. Reverse-osmosis desalination illustrates the challenge: achieving a few litres per square metre per hour per bar of throughput across membranes with subnanometre pores requires tens to hundreds of thousands of square metres of active area in industrial plants, each membrane element built from polymer structures whose selectivity resides at angstrom scales. The problem is therefore to bridge two extremes, translating angstrom-scale selectivity into infrastructure-scale membrane performance[95]. Nanofluidic studies have shown promise to bridge this gap. Investigations of electrokinetics in model nanochannels[128] inspired the design of bio-based osmotic energy membranes that outperform conventional polymers. These concepts have now been implemented in pilot-scale installations that harvest blue energy from river–sea salinity gradients, for example, the *Sweetch Energy* pilot plant on the Rhône river (2025). Another example is the development of graphene-oxide laminates for ultrafast water pervaporation[85], a concept now being explored industrially, including by the *Carlsberg Group* for dehydration applications.

Nanoscale water offers many assets. Scaling up, however, requires an intermediate mesoscopic description where membranes are treated as networks of pores with emergent properties that are not simple sums of single-pore behaviours[129]. Concepts from percolation, effective-medium theories, and non-equilibrium thermodynamics have long been used to describe porous media, but they rarely incorporate the non-classical behaviours of nanoscale water—such as dielectric anomalies, quantum-origin friction, matrix fluctuations, memristive ionic transport and, maybe most importantly, non-linear response[95,125,129].

At the same time, theoretical models typically describe confined water in idealised systems with simplified structural assumptions, whereas real nanopores are more complicated and exceptionally hard to control experimentally at the molecular scale under ambient conditions. There will for example inevitably be hydrocarbon contamination from air or fouling from solutions, irregularities and defects, and even finite elasticity of confinement, all of which may affect the outcome. Developing mesoscale models that explicitly encode nanofluidic laws while accounting for this structural complexity is therefore a crucial step if we are to design membranes whose collective properties transcend the limitations of today's permeability–selectivity trade-offs[125,129].

**Digital water gets wet**

Classical and ab initio molecular dynamics have proven useful for simulating interfacial and nanoconfined water[116,130–133]. However, the emergence of ML interatomic potentials has become transformative. Simulations at density functional theory or higher-level accuracy[134–137] at costs comparable to classical force fields are now possible. Architectures such as MACE[138] now achieve exceptionally small errors (meV-per-atom energy errors and tens of meV-per-angstrom force errors) on training sets that include bulk, interfacial, and confined water configurations. These approaches now enable simulations with thousands of atoms over nanosecond timescales on an almost routine basis[139,140,138]. They have been used to map phase diagrams of nanoconfined water[29,30], to compute friction coefficients[76,141], to explore processes such as hydroxide adsorption[10,76], proton transfer[142], and more. This is promising, however, a note of caution is necessary as challenges remain, with e.g., the robust treatment of long-range electrostatics[143], non-adiabatic[144] and spin effects. In addition, routinely going "beyond" the accuracy offered by density functional theory is challenging, and although ML simulations are much more efficient than AIMD, the system sizes and timescales accessible are still small and short compared to experiment.

In parallel, experimental techniques—HD-SFG, ultrafast 2D vibrational spectroscopy, NV-centre-based quantum sensing—provide increasingly detailed, quantitative benchmarks for this "digital water". A compelling example is the recent combined study of hydroxide adsorption on hBN[10], where ML-molecular dynamics and HD-SFG spectroscopy revealed two nearly isoenergetic adsorption configurations separated by a low barrier, whose interconversion reproduces the pH-dependent charging and spectroscopic signatures observed experimentally. Achieving such consistency across theory and experiment required not only accurate ML potentials but also a back-and-forth validation of ML predictions against experimental data. To some extent, an "AI moment" has arrived and the combination of ML forcefields and automated (agentic) approaches is likely to become a central "technology" of nanoscale water research.

Nanoscale water provides a demanding testing ground for such AI-augmented, experiment-anchored science. Dielectric anisotropy in angstrom slits, singular ion statistics in nanochannels, ultra-fast or ultra-slow flows, and confinement-induced phase transitions, to cite a few, are all sensitive to subtle free-energy differences and thus provide stringent tests for ML models. At the same time, the technological relevance of water in energy, desalination, and health means that better predictions and control of nanoscale water could rapidly translate into new materials and devices.

**Outlook: crossing the molecular frontier**

Nanoscale water occupies the frontier where continuum descriptions meet the molecular and sometimes quantum nature of matter. Classical concepts such as dielectric constants, surface friction, ionic mobilities and reaction rates remain indispensable at mesoscopic scales, but near surfaces and under molecular-scale confinement they must be generalised to account for anisotropy, nonlocal response, fluctuations and coupling to the electronic, vibrational and chemical degrees of freedom of the confining walls.

Three intertwined directions seem especially important. First, experiments should be pushed further, aiming to resolve how water and ions fluctuate, correlate, couple to walls and participate in chemical reactions in extreme 1D and 2D confinement. Second, AI-augmented modelling should be developed in close connection with experiment, not as a black box but as a way to generate predictions that can be challenged and refined against real confined and interfacial water. Third, molecular insight must be translated into mesoscale theories that can feed into the design of membranes, porous networks and devices, helping to bridge the gap between angstrom-scale selectivity and infrastructure-scale performance.

As these efforts mature, nanoscale water should become more than a collection of anomalies at the limits of continuum theory. It will serve both as a testing ground for physical principles at the boundary between continuum, molecular and quantum descriptions, and as a basis for technologies spanning clean water, sustainable energy and bio-inspired information processing. Because nanoconfined and interfacial water is ubiquitous in nature and technology, its predictive molecular understanding should have consequences well beyond nanofluidics, informing technologies that already rely on water and ions and enabling new ones. The deeper goal is not simply to catalogue nanoscale anomalies, but to make the behaviour of water predictable and controllable at the molecular scale. This could turn one of the most familiar substances into a material for design, much as earlier technologies were transformed by learning to control metals, silicon and polymers.